\documentclass[12pt]{article}
\usepackage{amsmath,slashed}
\usepackage{amssymb}
\usepackage{anysize}
\usepackage{graphicx}
\usepackage{bbold}
\usepackage{multicol}
\usepackage{dsfont}
\usepackage{setspace}
\usepackage{url}
\usepackage{hyperref}
\usepackage{multirow}
\usepackage{pbox}
\usepackage{xcolor}

\makeatletter\@addtoreset{equation}{section}\makeatother

\newcommand{\be}{\begin{equation}}
\newcommand{\ee}{\end{equation}}
\newcommand{\ben}{\begin{equation*}}
\newcommand{\een}{\end{equation*}}
\newcommand{\bea}{\begin{eqnarray}}
\newcommand{\eea}{\end{eqnarray}}
\newcommand{\nn}{\nonumber}
\newcommand{\bean}{\begin{eqnarray*}}
\newcommand{\eean}{\end{eqnarray*}}
\newcommand{\dd}{\mathrm{d}}
\newcommand{\p}{\partial}
\newcommand{\rt}{\right}
\newcommand{\lt}{\left}

\newcommand{\ba}{\begin{array}}
\newcommand{\ea}{\end{array}}

\newcommand{\mc}{\mathcal}

\newcommand{\dg}{\dagger}

\newcommand{\ov}{\overline}

\newcommand{\address}[1]{\vbox{\center\em#1}}

\newcommand{\ga}{\gamma}
\newcommand{\Ga}{\Gamma}
\newcommand{\de}{\delta}

\newcommand{\ep}{\epsilon}
\newcommand{\vep}{\varepsilon}

\newcommand{\Lm}{\Lambda}

\newcommand{\om}{\omega}
\newcommand{\Om}{\Omega}

\newcommand{\fs}{\!\!\!/}
\newcommand{\D}{\nabla\!\fs}

\newcommand{\dsT}{\mathds{T}}
\newcommand{\cA}{\mathcal A}
\newcommand{\cF}{\mathcal F}
\newcommand{\cR}{\mathcal R}

\begin{document}
\bibliographystyle{utphys}
\begin{titlepage}

\hfill{}
\\
\vspace{10mm}

\center{\LARGE K\"ahler Potential and Ambiguities in 4d  ${\cal N}=2$ SCFTs}
\vspace{15mm}

\begin{center}
\renewcommand{\thefootnote}{$\alph{footnote}$}
{Jaume Gomis\,\footnote{\href{mailto:jgomis@perimeterinstitute.ca}{\tt jgomis@perimeterinstitute.ca}} and Nafiz Ishtiaque\footnote{\href{mailto:nishtiaque@perimeterinstitute.ca}{\tt nishtiaque@perimeterinstitute.ca}}   
}

\vspace{18mm}

\address{${}^{a,b}$Perimeter Institute for Theoretical Physics,\\
Waterloo, Ontario, N2L 2Y5, Canada}
\address{${}^{a}$Department of Physics, University of Waterloo\\
Waterloo, Ontario N2L 3G1, Canada}

\renewcommand{\thefootnote}{\arabic{footnote}}
\setcounter{footnote}{0}

\end{center}

\vspace{25mm}

\abstract{
\medskip\medskip
\normalsize{
\noindent
The partition function of four-dimensional $\mc N=2$  superconformal field theories on $S^4$ computes the exact K\"ahler potential  on the space of exactly marginal couplings~\cite{Gerchkovitz:2014gta}. We present a new elementary proof of this result using   supersymmetry Ward
	identities. The partition function is a section rather than a function, and is subject to ambiguities  coming from  K\"ahler transformations acting on the K\"ahler potential. This ambiguity  is realized by  a  
	local supergravity counterterm in the underlying SCFT. 
 We provide an explicit construction of the K\"ahler ambiguity counterterm in  the four dimensional ${\cal N}=2$ off-shell  supergravity theory that admits $S^4$ as a supersymmetric background.
}
}

\vspace{10mm}

\noindent
\vfill

\end{titlepage}

\section{Introduction} 
\label{sec:intro}

Recent years have witnessed remarkable progress in obtaining  the exact   partition function of supersymmetric field theories in various background geometries. When the geometry  is $S^1\times {\cal M}_{d-1}$  the partition function admits a standard Hilbert space   interpretation as a supertrace  over the  states   of the theory on  ${\cal M} _{d-1}$. In other   geometries, such as on a sphere  $S^d$, 
 the physical interpretation of the partition function   must be sought.

In~\cite{Gerchkovitz:2014gta}  it has been shown  that the partition function of 4d ${\cal N}=2$ superconformal field theories (SCFTs) on $S^4$ computes the exact K\"ahler potential $K$ on the space of exactly marginal couplings, also referred to as the conformal manifold. This result was proven both by using supersymmetric localization~\cite{Pestun:2007rz} and by conformal dimension regularization   on $S^4$, and  extends the proof in~\cite{Gomis:2012wy}   that    the $S^2$ partition function of 2d ${\cal N}=(2,2)$ SCFTs computes the exact K\"ahler potential
  on the conformal manifold, as conjectured by~\cite{Jockers:2012dk}  based on the exact formulae   in~\cite{Benini:2012ui,Doroud:2012xw} (see also~\cite{Gomis:2012wy,Doroud:2013pka}). In detail,~\cite{Gerchkovitz:2014gta}  demonstrated that  
  \be
Z_{S^4}=e^{K/12}\,.
\label{spherekahler}
\ee
These identifications  provide a physical and geometrical interpretation of  the sphere partition function of 4d ${\cal N}=2$ and 2d ${\cal N}=(2,2)$  SCFTs.  These results also provide a computational pathway for obtaining the exact metric in the conformal manifold, which are interesting new    observables in these theories, acted on by dualities (see e.g. recent work~\cite{Baggio:2014ioa}\cite{Baggio:2014sna}).

 Here  we present an   elementary proof of the formula  (\ref{spherekahler}) using   supersymmetry Ward identities. This new proof  does not require localization nor that the 4d ${\cal N}=2$ SCFT admits a Lagrangian description. By virtue of the relation (\ref{spherekahler}) identifying the $S^4$  partition function with the K\"ahler potential $K$ on the conformal manifold, it follows that the partition function is subject to  the K\"ahler ambiguity transformations 
   \be
  K(\tau,\bar \tau)\rightarrow K(\tau,\bar \tau)+{\cal F}(\tau)+\bar{\cal F}(\bar\tau)\,,
  \label{kahler}
  \ee
  where ${\cal F}$ is an arbitrary    holomorphic function and $\tau$ are holomorphic coordinates on the conformal manifold. 
  This ambiguity implies that the partition function is a section over the space of exactly marginal couplings.

We also give the microscopic realization of the K\"ahler ambiguity (\ref{kahler}) by constructing the local supergravity counterterm  in 4d ${\cal N}=2$ off-shell supergravity 
 that when evaluated on the supersymmetric $S^4$ background yields (\ref{kahler}). 
 This  is the 4d counterpart  of the K\"ahler ambiguity counterterm for 2d ${\cal N}=(2,2)$ SCFTs constructed in~\cite{Gerchkovitz:2014gta}.
     
     The plan is as follows. In section \ref{sec:susyward} we use supersymmetry Ward identities to show that the $S^4$ partition function of  4d ${\cal N}=2$ SCFTs computes the K\"ahler potential in the conformal manifold. In section \ref{kahlersugra} we identify the off-shell 4d  ${\cal N}=2$ Poincar\'e supergravity theory in which the  $S^4$ is a supersymmetric background. In section \ref{kahlersugrab} we construct the supergravity invariant in the  relevant Poincar\'e supergravity theory  that once evaluated on  $S^4$  provides a first principles realization of the K\"ahler transformation (\ref{kahler}).

 \section{K\"ahler Potential from \texorpdfstring{$S^4$}{Lg} Partition Function}
\label{sec:susyward}

 An exactly marginal operator  in a  four dimensional ${\cal N}=2$ SCFT   is a scalar  operator 
of   dimension four   which is a superconformal descendant of 
a scalar chiral primary operator of $U(1)_R$ charge $w=2$.
An ${\cal N}=2$ SCFT can be deformed while preserving
superconformal invariance  by\footnote{We use the same conventions as in~\cite{Gerchkovitz:2014gta}.}
\be
{1\over \pi^2}\int d^4x\, \sum_I \left( \tau_I O_I+\bar \tau_{\bar I} \bar O_{\bar I}\right)\,.
\ee
The exactly marginal couplings $\tau_I$ are holomorphic   coordinates in the space of exactly marginal deformations, known as  the conformal manifold.
The canonical metric in the conformal manifold $g_{I\bar J}$ is the Zamolodchikov metric 
\be
\langle O_I(x) \bar O_{\bar J}(0)\rangle={g_{I\bar J}\over x^8}\,,
\ee
which in four dimensional ${\cal N}=2$ SCFTs is K\"ahler, that is
 \be
 g_{I\bar J}={\partial\over \partial \tau_I}{\partial\over \partial \bar\tau_{\bar J}}K(\tau,\bar \tau)\equiv\partial_I\partial_{\bar J}K(\tau,\bar\tau)\,.
 \label{kahlerdef}
\ee

An ${\cal N}=2$ SCFT  can be canonically placed on $S^4$ by the stereographic projection. The ${\cal N}=2$  superconformal transformations on $S^4$ are parametrized by
chiral conformal Killing spinors $\epsilon^i$ and $\epsilon_i$ of opposite chirality  transforming as doublets of the  $SU(2)_R$ R-symmetry, which obey\footnote{$\Gamma^a$ denotes tangent space gamma matrices while $\gamma^m=e_a^m\Gamma^a$  denotes curved space ones.}
 
\be
\nabla_m \epsilon^i=\gamma_m\eta^i\qquad \nabla_m \epsilon_i=\gamma_m\eta_i\,,
\label{cks}
\ee
 so that $\eta^i={1\over 4}\D\epsilon^i$ and $\eta_i={1\over 4}\D\epsilon_i$.
 
An exactly marginal operator in an ${\cal N}=2$ SCFT can be represented as the top component of a four dimensional ${\cal N}=2$ {\it chiral multiplet} of R-charge $w=2$,   whose bottom component  realizes the parent chiral primary operator. 
The holomorphic coordinates on  the conformal manifold can be promoted to supersymmetric background chiral superfields with vanishing  R-charge $w=0$.
The ${\cal N}=2$ superconformal transformations of a chiral multiplet with R-charge $w$ on $S^4$ are given by~\cite{Breitenlohner:1980ej} (we use~\cite{parisLect}):\footnote{Throughout a barred spinor is $\bar\lambda=\lambda^T {\cal C}$, where ${\cal C}$ is the charge conjugation matrix.}

	\begin{align}
	 \de  A &= \frac{1}{2} \ov\ep^i \Psi_i \nn\\
		\de \Psi_i &= \D(A\ep_i) + \frac{1}{2} B_{ij} \ep^j + \frac{1}{4} \Ga^{ab} F_{ab}^- \vep_{ij} \ep^j 
			+ (2w-4) A \eta_i \nn\\
		\de  B_{ij} &= \ov\ep_{(i} \D\, \Psi_{j)} - \ov\ep^k \Lm_{(i} \vep_{j)k} + 2(1-w) \ov\eta_{(i} \Psi_{j)} \nn\\
		\de F_{ab}^- &= \frac{1}{4} \vep^{ij} \ov\ep_i \D\, \Ga_{ab} \Psi_j + \frac{1}{4} \ov\ep^i \Ga_{ab} \Lm_i 
			- \frac{1}{2}(1+w) \vep^{ij} \ov\eta_i \Ga_{ab} \Psi_j \nn\\
		\de \Lm_i &= -\frac{1}{4} \Ga^{ab} \D\, (F_{ab}^- \ep_i) - \frac{1}{2} \D B_{ij} \vep^{jk} \ep_k + \frac{1}{2} C \vep_{ij} \ep^j -(1+w) B_{ij} \vep^{jk} \eta_k + \frac{1}{2} (3-w) \Ga^{ab} F_{ab}^- \eta_i \nn\\ 
				\de C &= -\nabla_m (\vep^{ij} \ov\ep_i \ga^m \Lm_j) + (2w-4) \vep^{ij} \ov\eta_i \Lm_j\,,
			\label{dC}
	\end{align}
	where in Euclidean signature  $F_{ab}^-$ is a self-dual rank-two tensor. Indeed, for $w=2$, the integrated top component is superconformal invariant and we have the identification
\be
C_I=O_I\qquad \hbox{for}~w=2\,.
\ee
For $w=0$, an arbitrary covariantly constant background value for the  bottom component of the chiral multiplet\footnote{All other components in multiplet must vanish.} is superconformal invariant,
 and    serves as the spurion field  for the holomorphic coordinates on the conformal manifold
\be
 A_I  =\tau _I\qquad \hbox{for}~w=0\,.
 \label{spurii}
\ee
 We denote by ${\cal A}_I$ the chiral multiplets to which the coordinates in the conformal manifold have been promoted.

Consider now the SCFT partition function on $S^4$ as a function of the exactly marginal  couplings $Z_{S^4}(\tau,\bar \tau)$.  The   second derivative 
\be
\partial_I \partial_{\bar J} \log Z_{S^4}={1\over \pi^4} \left\langle \int_{S^4} d^4x\sqrt{g}\, C_I(x)\   \int_{S^4} d^4y \sqrt{g}\,  \bar C_{\bar J}(y)\right\rangle\, 
\label{intetwo}
\ee
  is the {\it integrated} connected two-point function of  exactly marginal operators. This correlator is ultraviolet divergent, divergences arising when the  operators collide. These ultraviolet divergences can be regularized by introducing a massive deformation. Regulating   divergences in 
a supersymmetric manner leads us to consider the $OSp(2|4)$  massive subalgebra of the ${\cal N}=2$ superconformal algebra on $S^4$, which is the supersymmetry algebra of an arbitrary massive four dimensional ${\cal N}=2$ theory on $S^4$.

The  $OSp(2|4)$ massive subalgebra on $S^4$ is generated by supercharges that anticommute to the $SO(5)$ isometries of $S^4$ and an $SO(2)_R\subset SU(2)_R$ R-symmetry. Conformal generators and $U(1)_R$ are projected out. The $OSp(2|4)$ transformations are generated by   Killing spinors which   obey
\be
\nabla_m \chi^j = \frac{i}{2r} \ga_m \chi^j\,,
\label{killSPH}
\ee
where 
\be
 \chi^j = \ep^j + \tau_1^{jk} \ep_k 
\ee
so that\footnote{$P_{L}$ and $P_{R}$ are the spinor chirality projectors: $P_{L}^2=P_{L}$, $P_{R}^2=P_R$ and $P_L+P_R=1$. The Killing spinors obey
 $P_L\epsilon^i=\epsilon^i$ and $P_R\epsilon_i=\epsilon_i$.}
\be
\epsilon^i=\chi^i_L\qquad \epsilon_i=\tau_{1ij} \chi^j_R
\ee
and $\tau_p^{jk}=(i \sigma_3,-1,-i \sigma_1)=({\tau_p}_{jk})^*$, where    $\sigma_p$ are the Pauli matrices.
In stereographic coordinates, where $ds^2={1\over \left(1+ {x^2\over 4r^2}\right)^2} dx_m dx^m$,  we have
\be
\chi^j={1\over \sqrt{1+{x^2\over 4r^2}}} \left(1+ {i \over 2r} x_m\Gamma^m\right) \chi_0^j\,.
\label{killsph}
\ee
 The constant spinors   $\chi_0^j$ parametrize the transformations of the eight supercharges in $OSp(2|4)$. If these parameters are chiral
 \be
 P_L \chi^j_0=0\,,
 \label{chiralconst}
 \ee
 the corresponding spinors generate an  $OSp(2|2)$ subalgebra $OSp(2|4)$.  The chiral components of these spinors $\chi_L^j$ and $\chi_R^j$ 
 \be
 \chi^j_L=P_L \chi^j={i/2r\over \sqrt{1+{x^2\over 4r^2}}}   {x_m} \Gamma^m   \chi_{0R}^j\qquad 
\chi^j_R=P_R \chi^j={1\over \sqrt{1+{x^2\over 4r^2}}}    \chi_{0R}^j\,, 
\label{osp2}
 \ee
 vanish at the North and the South poles of the sphere respectively.  If the parameters are further constrained by 
  \be
  \chi^i_0 = \tau_1^{ij} \vep_{jk} \Ga_{1}\Ga_{2} \chi^k_0\,, \label{su11}
\ee
  the   corresponding spinors generate  a  further $SU(1|1)$   subalgebra 
  \be
  Q^2=J+R
  \ee
   of $OSp(2|2)\subset OSp(2|4)$, where $J=J_{12}+J_{34}$ is a self-dual rotation on $S^4$ and $R$ is the   $SO(2)_R\subset SU(2)_R$ R-symmetry.

Our strategy  is to first prove that the {\it integrated} top component of the chiral multiplet  in (\ref{intetwo}) can be written as  an $SU(1|1)\subset OSp(2|4)$ supersymmetry transformation $\delta$ everywhere except 
 at  the North pole of $S^4$, where the corresponding Killing spinor  vanishes.  The proof is completed by showing  that the correlator of   the integrated top component $C$ with an arbitrary   operator ${\cal O}$  invariant under the $SU(1|1)$ supersymmetry transformation $\delta$  reduces to the correlator  of   the bottom component $A$ at  the North pole with  ${\cal O}$. In detail
 \be
 \left\langle \int_{S^4} d^4x\sqrt{g}\, C(x)\, {\cal O} \right\rangle= 32 \pi^2r^2 \left\langle A(N)\, {\cal O} \right\rangle  \,.
 \label{desired}
 \ee
 
 The supersymmetry transformation of the fermions in a chiral multiplet with R-charge $w=2$ can be written as (\ref{dC})\footnote{$\vec B = (B_1, B_2, B_3)$ such that 
$B_{ij} = \vec B \cdot \vec\tau_{ij} = \sum_p B_p \tau_{pij}$.}
\begin{subequations}\begin{align}
	\de \Psi_i =&\; \tau_{1ij} \slashed\nabla (A\chi^j_R) + \frac{1}{2} \vec B \cdot \vec\tau_{ij} \chi^j_L + \frac{1}{4} \Ga^{ab} F_{ab}^- \vep_{ij} \chi^j_L\label{susyf} \\
	\de \Lm_i =& -\frac{1}{4} \Ga^{ab} \slashed\nabla F_{ab}^- \tau_{1ij} \chi^j_R - \frac{i}{4r} \Ga^{ab} F_{ab}^- \tau_{1ij} \chi^j_L
		+ \frac{1}{2} C \vep_{ij} \chi^j_L \nn\\
		& \label{susyF} - \frac{1}{2} \slashed\nabla \vec B \cdot \vec \tau_{ij} \tau_1^{jk} \vep_{kl} \chi^l_R - \frac{3i}{2r} \vec B \cdot \vec\tau_{ij} \tau_1^{jk} \vep_{kl} \chi^l_L
\end{align}\end{subequations}
Using the $SU(1|1)$ supersymmetry transformation $\delta$ obtained by imposing the constraints (\ref{chiralconst}) and (\ref{su11}) on the Killing spinors, we get after multiplying
(\ref{susyf}) by $\tau_{2ij} \tau_1^{jk}{\chi^i_L}^\dg$ and (\ref{susyF}) by $\tau_{2ij} \vep^{jk} {\chi^i_L}^\dg$   that
 \begin{subequations}\begin{align}
	B_1 =& -\de\lt(\frac{{\chi^i_L}^\dg \Psi_k}{\Vert\chi_L\Vert^2}\rt) \tau_{2ij} \tau_1^{jk} + \frac{{\chi_L^i}^\dg}{\Vert\chi_L\Vert^2}
    	\slashed\nabla(A\chi^j_R) \tau_{2ij} - \frac{1}{4} \frac{{\chi^i_L}^\dg \Ga^{ab} \chi^j_L}{\Vert\chi_L\Vert^2} \tau_{3ij} F_{ab}^- \label{B1Ca}\\
    C =& -\de\lt(\frac{{\chi^i_L}^\dg \Lm_k}{\Vert \chi_L\Vert^2}\rt) \tau_{2ij} \vep^{jk} - \frac{1}{2} \frac{{\chi^i_L}^\dg \ga^m \chi^j_R}
    		{\Vert\chi_L\Vert^2} \tau_{2ij} \nabla_m B_1 + \frac{3i}{r} B_1 \nn\\
    		& + \frac{1}{4} \frac{{\chi^i_L}^\dg \Ga^{ab} 
    		\ga^m \chi^j_R}{\Vert \chi_L\Vert^2} \tau_{3ij} \nabla_m F_{ab}^- + \frac{i}{4r} \frac{{\chi^i_L}^\dg \Ga^{ab} \chi^j_L}{\Vert\chi_L\Vert^2} \tau_{3ij} F_{ab}^- \label{B1Cb}
\end{align}\label{B1C}\end{subequations}
where we have used that for the $SU(1|1)$ Killing spinors $\Vert\chi_L\Vert^2 := \Vert\chi^1_L\Vert^2 = \Vert\chi^2_L\Vert^2$,   
where $\Vert\lambda \Vert^2=\lambda^\dagger\lambda$. The terms proportional to $F_{ab}^-$ and
$\nabla_\mu F_{ab}^-$ in (\ref{susyf}) (\ref{susyF}) also vanish. Their coefficients are anti-self-dual in the tangent space indices since
\be
{\chi^i_L}^\dg \Ga^{ab} \ga^{(r)} 
\chi^j_{L/R} = {\chi^i_L}^\dg \Ga_* \Ga^{ab}\ga^{(r)} \chi^j_{L/R} = -\frac{1}{2}\vep^{ab}_{\;\;\;\,cd} {\chi^i_L}^\dg\Ga^{cd}\ga^{(r)} \chi^j_{L/R}\,,
\ee
where $\ga^{(r)}$ is the product of $r$ distinct gamma matrices. Since $F_{ab}^-$ is self-dual in Euclidean signature, all the
terms involving $F_{ab}^-$ vanish.
	We can eliminate $B_1$ from (\ref{B1Cb}) by using (\ref{B1Ca}), which yields 
	 \begin{align}
		C =&\;  - \frac{1}{2} \frac{{\chi^i_L}^\dg}{\Vert\chi_L\Vert^2} \slashed\nabla \lt( \lt[ \frac{{\chi^k_L}^\dg}{\Vert\chi_L\Vert^2} \slashed\nabla (A \chi^l_R) \rt] \chi^j_R\rt) \tau_{2ij} \tau_{2kl}
				+ \frac{i}{r} \frac{{\chi^i_L}^\dg}{\Vert\chi_L\Vert^2} \slashed\nabla (A\chi^j_R) \tau_{2ij} \nn\\
			& +\de\lt(\Xi(\Lm_i, \Psi_i, \chi^i)\rt)\,, \label{C}
    \end{align}
	where, for brevity, we have defined 
	\be 
		\Xi(\Lm_i, \Psi_i, \chi^i) := -\frac{{\chi^i_L}^\dg \Lm_k}{\Vert \chi_L\Vert^2} \tau_{2ij} \vep^{jk} + \frac{1}{2} \frac{{\chi^i_L}^\dg \ga^m \chi^j_R}{\Vert\chi_L\Vert^2}
				\nabla_m \lt( \frac{{\chi^k_L}^\dg \Psi_t}{\Vert\chi_L\Vert^2} \rt) \tau_{2ij} \tau_{2kl} \tau_1^{lt}
				- \frac{3i}{r} \frac{{\chi^i_L}^\dg \Psi_k}{\Vert\chi_L\Vert^2} \tau_{2ij} \tau_1^{jk}\,.
	\ee
We now show that the sum of the terms in (\ref{C}) involving $A$ are a total derivative. 

	For any $OSp(2|2)$ supersymmetry parameter $\chi^j$ and any scalar quantity $X$ we have that\footnote{By using that  $\nabla_m \chi_L^\dg = -\frac{i}{2r} \chi_R^\dg \ga_m$.}
	\be 
		\frac{{\chi^j_L}^\dg}{\Vert \chi^j_L \Vert^2} \D \left(X \chi^j_R\right) 
		= \nabla_m \lt( \frac{{\chi^j_L}^\dg \ga^m \chi^j_R}{\Vert \chi^j_L \Vert^2} X \rt) + \frac{4irX}{x^2}\,.	\label{D2div1}
	\ee
 Using this, the top component $C$ of a chiral multiplet with $w=2$ can be written locally as the sum of an $SU(1|1)$ supersymmetry transformation $\delta$ and total derivatives 
  \begin{align}
		C =&\; \de\lt(\Xi(\Lm_i, \Psi_i, \chi^i)\rt) - \frac{1}{2} \nabla_m \lt( \frac{{\chi^i_L}^\dg \ga^m \chi^j_R}{\Vert \chi_L \Vert^2}
			\nabla_n \lt[\frac{{\chi^k_L}^\dg \ga^n \chi^l_R}{\Vert \chi_L \Vert^2} A \rt]\rt) \tau_{2ij} \tau_{2kl} \nn\\
			& +8ir \nabla_m\lt( \frac{{\chi^i_L}^\dg \ga^m \chi^j_R A}{\Vert \chi_L\Vert^2 x^2}\rt) \tau_{2ij}
			+ \frac{i}{r} \nabla_m \lt( \frac{{\chi^i_L}^\dg \ga^m \chi^j_R}{\Vert \chi_L \Vert^2}A\rt) \tau_{2ij}\,.
		 \label{finn}
	\end{align}
This formula fails  at the North pole, where 
 $\Vert \chi^1_L \Vert=\Vert \chi^2_L \Vert=0$ and $\Xi$ diverges. Therefore the integrated top component is  non-trivial in correlation functions,  as it is not 
 supersymmetry-exact globally, but the entire contribution localizes to the North pole, just as in the analysis of 2d ${\cal N}=(2,2)$ SCFTs in~\cite{Gerchkovitz:2014gta}.\footnote{We note that had we assumed that the partition function can be regulated while preserving full ${\cal N}=2$ superconformal invariance, we would have concluded that the 
 partition function is independent of the moduli, as the top component $C$  is globally superconformal-exact.}
  
Let us consider the integrated correlator  with an operator ${\cal O}$ obeying
  $\delta{\cal O}=0$
 \be
  \left\langle \int_{S^4} d^4x\sqrt{g}\, C(x)\, {\cal O}  \right\rangle=\lim_{R\rightarrow 0} \left[\left\langle \int_{S^4 \backslash B^4_R} d^4x\sqrt{g}\, C(x)\, {\cal O}  \right\rangle+    \left\langle \int_{B^4_R} d^4x\sqrt{g}\, C(x)\, {\cal O}  \right\rangle\right]\,.
 \ee
We have divided $S^4$ into two-regions: a four-dimensional ball $B^4_R$ of radius $R$ around the North pole and its complement $S^4 \backslash B^4_R$. In the $R\rightarrow 0$ limit the ball contribution vanishes\footnote{The $R^4$ measure factor suppresses the ball contribution in the $R\rightarrow 0$ limit.} and we are left with
 \be
 \lim_{R\rightarrow 0} \left\langle \int_{S^4 \backslash B^4_R} d^4x\sqrt{g}\, C(x)\, {\cal O}  \right\rangle\,.
 \label{integraregio}
 \ee
 Using (\ref{finn}), which is valid in $S^4 \backslash B^4_R$, and   $\delta\Phi=0$, we can replace $C$ by the last three terms in (\ref{finn}),  which inside (\ref{integraregio}) can be written as an integral over the three-sphere $S^3_R$ of radius $R$ at the boundary of $S^4 \backslash B^4_R$. For any $OSp(2|2)$ Killing spinor $\chi^j$ (\ref{osp2}),  we have that in the $R \to 0$
 limit 
	\be 
		\chi^i_L(R) \sim   O(R),\; \chi^i_R(R) \sim  O(1) \Rightarrow \frac{{\chi^i_L}^\dg \ga^\mu \chi^i_R}{\Vert\chi_L\Vert^2} \sim    O\lt(\frac{1}{R}\rt)\,.\label{r2}
	\ee
Therefore, 	a simple scaling argument shows that the last term in (\ref{finn}) cannot compensate for the $R^3$ measure factor coming from   $S^3_R$   and gives a vanishing contribution in the $R\rightarrow 0$ limit. Therefore, we have shown that in the presence of   $\delta$-closed operators 
\begin{align}
  \int_{S^4} d^4x\sqrt{g}\, C(x)  =& \lim_{R\rightarrow 0}  \int_{S^4 \backslash B^4_R} d^4x\sqrt{g}\, C(x)\cr 
  	=& -\frac{1}{2}\, \lim_{R\rightarrow 0}  \int_{S^4 \backslash B^4_R} \dd^4 x\, \p_m \lt( \frac{{\chi^i_L}^\dg \ga^m \chi^j_R}{\Vert \chi_L \Vert^2}
			\p_n \lt[\frac{{\chi^k_L}^\dg \ga^n \chi^l_R}{\Vert \chi_L \Vert^2} A(x) \sqrt{g} \rt]\rt) \tau_{2ij} \tau_{2kl}   \nn\\
			&\label{int1} +8ir \lim_{R\rightarrow 0}  \int_{S^4 \backslash B^4_R} \dd^4 x\, \p_m
			\lt( \frac{{\chi^i_L}^\dg \ga^m \chi^j_R}{\Vert \chi_L \Vert^2 x^2} A(x) \sqrt{g}\rt) \tau_{2ij}\,. 
 \end{align}
In the limit $R\rightarrow 0$  we can replace the bottom component $A(x)$ by its value at the North pole $A(N)$, as higher order terms in the expansion in $R$ vanish in the limit, and  using Stoke's theorem
\be 
  \int_{S^4} d^4x\sqrt{g}\, C(x) = \lim_{R\rightarrow 0}  \int_{S^3_R} V \cdot \hat\eta\,, \label{int2}
	\ee
	where 
	\be
		V^m :=  -\frac{1}{2}\frac{{\chi^i_L}^\dg \ga^m \chi^j_R}{\Vert \chi_L \Vert^2} \p_n \lt(\frac{{\chi^k_L}^\dg \ga^n \chi^l_R}{\Vert \chi_L \Vert^2} \sqrt{g} \rt)  A(N) \tau_{2ij} \tau_{2kl} 
		+ 8ir \frac{{\chi^i_L}^\dg \ga^m \chi^j_R }{\Vert \chi_L \Vert^2 x^2} A(N) \sqrt{g} \tau_{2ij}\,,
	\ee
	and $\hat \eta$ is the unit vector towards the North pole of $S^4$ along the radial direction.\footnote{The unit radial vector  in cartesian coordinates is given by 
$\hat\eta^a = -\frac{x^a}{\sqrt{x^2}}$.}
Going to spherical coordinates, where $R$ is the radial coordinate, we find that 
	\be 
		V \cdot \hat\eta = \frac{512 A(N) r^6 (R^2 - 2r^2)}{R^3 (R^2 + 4r^2)^3} + \frac{2048 A(N) r^8}{R^3 (R^2 + 4r^2)^3} = \frac{512 A(N) r^6 (R^2 + 2r^2)}{R^3 (R^2 + 4r^2)^3}\,.
	\ee
	The integration in (\ref{int2}) is over  $S^3_R$,  therefore
	\be
		 \int_{S^3_R} V \cdot \hat\eta = \frac{512 A(N) r^6 (R^2 + 2r^2)}{R^3 (R^2 + 4r^2)^3} 2 \pi^2 R^3\,,
	\ee
and
	\begin{align}
		& \lim_{R \to 0} \frac{512 A(N) r^6 (R^2 + 2r^2)}{R^3 (R^2 + 4r^2)^3} 2 \pi^2 R^3 = 32 A(N) \pi^2 r^2\,.
	\end{align}
	This yields the desired  formula
	\be
 \left\langle \int_{S^4} d^4x\sqrt{g}\, C(x)\, {\cal O} \right\rangle= 32 \pi^2r^2 \left\langle A(N)\, {\cal O}  \right\rangle  \,.
 \label{finC}
 \ee
	The integrated top component $C$ of a chiral multiplet is equivalent to inserting the bottom component $A$ at the North pole.
A very similar analysis yields 	   
 	\be
 \left\langle \int_{S^4} d^4x\sqrt{g}\, \bar C(x)\, {\cal O}  \right\rangle= 32 \pi^2r^2 \left\langle \bar A(S)\, {\cal O}  \right\rangle  \,.
  \label{finAC}
 \ee
  	The integrated top component $\bar C$ of an anti-chiral multiplet is equivalent to inserting the bottom component $\bar A$ at the South pole.

   We can now use (\ref{finC}) and (\ref{finAC}) to express the derivative of the partition function in (\ref{intetwo}) as an {\it unintegrated}  two-point function
\begin{align}
\partial_I \partial_{\bar J} \log Z_{S^4}&={1\over \pi^4} \left\langle \int_{S^4} d^4x\sqrt{g}\, C_I(x)\   \int_{S^4} d^4y \sqrt{g}\,  \bar C_{\bar J}(y)\right\rangle\,= \left(32 r^2\right)^2\left\langle A_I(N) \bar A_{\bar J}(S) \right\rangle\,.
\label{intetwofin}
\end{align}
 It follows from the first equation in (\ref{dC}) that the  correlator $\left\langle A_I(N) \bar A_{\bar J}(S) \right\rangle$ is $\delta$ invariant, since $SU(1|1)$ supersymmetry parameters $\epsilon^j$ and $\epsilon_j$ vanish at the North pole and South pole respectively, and therefore $\delta A_I(N)=\delta \bar A_{\bar J}(S)=0$.

Using the supersymmetry Ward identity    $\left\langle A_I(N) \bar A_{\bar J}(S) \right\rangle={r^4\over 48} \left\langle C_I(N) \bar C_{\bar J}(S) \right\rangle$~\cite{Gerchkovitz:2014gta}, that $\left\langle C_I(N) \bar C_{\bar J}(S) \right\rangle={1\over (2r)^8} g_{I\bar J}$ defines the Zamolodchikov metric $g_{I\bar J}$ and that the metric is K\"ahler (\ref{kahlerdef}) we arrive at 
\be
\partial_I \partial_{\bar J} \log Z_{S^4}={1\over 12}g_{I\bar J}={1\over 12}\partial_I\partial_{\bar J} K\,.
\ee
Therefore, the four sphere partition function of a four dimensional ${\cal N}=2$ SCFT computes the K\"ahler potential in the conformal manifold (\ref{spherekahler}), and
is subject to  K\"ahler transformation ambiguities (\ref{kahler}), which do not affect the Zamolodchikov metric.

	\section{Off-shell ${\cal N}=2$ Poincar\'e Supergravity for $S^4$}
	\label{kahlersugra}
	
	The  partition function of a  field theory in a curved geometry can be     ambiguous. These ambiguities are encoded
	in finite counterterms for the background fields that capture the background geometry and the parameters of the theory.  When the partition function of a supersymmetric theory can be regulated in a diffeomorphism invariant and supersymmetric manner,  the counterterms are {\it supergravity} invariants constructed out of the   supergravity multiplet  encoding the background geometry and the      supersymmetry multiplets  to which the other parameters of the theory can be promoted, since all parameters in a supersymmetric field theory can be promoted to background supermultiplets \cite{Seiberg:1993vc}.
	
	Constructing these supergravity invariants requires  identifying first the supergravity theory in which the curved  geometry over which the partition function is computed is a supersymmetric background. This can be analyzed   in the framework of off-shell supergravity~\cite{Festuccia:2011ws}. In this section we identify the four dimensional ${\cal N}=2$ off-shell Poincar\'e supergravity theory and the background fields in that supergravity multiplet that give  rise to the $OSp(2|4)$-invariant  four-sphere background geometry.

	A conceptual way of constructing off-shell Poincar\'e supergravity theories is to start with off-shell conformal supergravity and partially gauge fix the conformal symmetries down to   Poincar\'e   by adding compensating supermultiplets. Different choices of compensating multiplets give rise to different off-shell Poincar\'e supergravity theories, with different sets of auxiliary fields.\footnote{For instance, old and new minimal four dimensional ${\cal N}=1$ Poincar\'e supergravity arises from ${\cal N}=1$ conformal supergravity by using a compensating chiral and tensor multiplet respectively.} The Poincar\'e supersymmetry transformations of the gauge fixed theory are constructed by combining the Poincar\'e supersymmetry transformations in conformal supergravity with field dependent superconformal transformations that are needed to preserve the gauge choice.\footnote{We refer to  the~\cite{freedman12} for more background material and references, in particular for   4d ${\cal N}=2$ supergravity.}
	
	Our starting point is four dimensional ${\cal N}=2$ conformal supergravity~\cite{deWit:1980tn} (we refer to \cite{freedman12} for  more details). Off-shell ${\cal N}=2$ superconformal transformations are realized on the Weyl multiplet, whose independent fields are
	\begin{align}
	\text{bosonic:} &\, e^a_m, b_m,V_{m\, i}^{\ \ j}, A^R_m, T_{ab}^-, D\cr
	\text{fermionic:} &\, \psi_m^i, \chi^i\,.
	\label{weylm}
	\end{align}
The fields $e^a_m, b_m,V_{m\, i}^{\ \ j}, A^R_m, \psi_m^i$ are the gauge fields for translations, dilatations, $SU(2)_R$, $U(1)_R$ and Poincar\'e supersymmetry generators in the ${\cal N}=2$  superconformal algebra. The Weyl multiplet is completed by the bosonic auxiliary fields $T_{ab}^-$ and  $D$, and the fermionic auxiliary field $\chi^i$. In Euclidean signature  $T_{ab}^-$ is a self-dual rank-two tensor. The embedding of the $OSp(2|4)$-invariant  $S^4$ in conformal supergravity appeared in~\cite{Hama:2012bg}\cite{Klare:2013dka}.

Four dimensional ${\cal N}=2$ Poincar\'e supergravity~\cite{Ferrara:1976fu} contains a graviphoton gauge field $A_m$. This field   is furnished in the conformal approach by coupling
an abelian vector multiplet to the Weyl multiplet~\cite{deWit:1979ug}\cite{deWit:1979pq}. An ${\cal N}=2$ vector multiplet, also known as a restricted  chiral multiplet, is an ${\cal N}=2$ chiral multiplet (\ref{dC}) with    $w=1$ subject to  constraints, and consists of
\begin{align}
	\text{bosonic:} &\, X, A_m, Y_{ij} \cr
	\text{fermionic:} &\, \Omega_i
	\label{vector}
	\end{align}
 a complex scalar $X$, a gauge field $A_m$, a triplet of real auxiliary fields $Y_{ij}=Y_{ji}$ and gauginos $\Omega_i$.  The vielbein $e_m^a$ and gravitino $\psi_m^i$ of the Weyl multiplet and the gauge field $A_m$  
	in the vector multiplet complete the on-shell content of four dimensional ${\cal N}=2$ Poincar\'e supergravity multiplet. 
		
	The first step in constructing a Poincar\'e supergravity theory is to gauge fix special conformal transformations. This can be  accomplished by setting
\be
b_m=0\,.
\label{specialc}
\ee
In order to preserve this gauge, supersymmetry transformations must  be accompanied by a compensating special conformal transformation, which acts nontrivially on $b_m$. Fortunately, all elementary fields in conformal supergravity and all  fields in ${\cal N}=2$ matter multiplets transform trivially under special conformal transformations, and therefore the supersymmetry transformations of these fields are not modified by the gauge choice (\ref{specialc}).

	Dilatations and $U(1)_R$ are gauge fixed by setting~\cite{deWit:1980tn}
	\be
	X=\mu\,,
	\label{fixmass}
	\ee
	where $\mu$ is an arbitrary mass scale, while~\cite{deWit:1980tn} 
	\be
	\Omega_i=0
	\label{fixS}
	\ee
	fixes the special conformal supersymmetry transformations. Under   supersymmetry  \cite{freedman12} 
	\begin{align}
		\de X =\;& \frac{1}{2} \ov\ep^i \Om_i \\
		\de \Om_i =\;& \mc D\!\fs\, X \ep_i + \frac{1}{4} \Ga^{ab} \mc F_{ab} \vep_{ij} \ep^j + {Y_{ij}\over 2} \ep^j + 2X\eta_i \,,
 \end{align}
 where $\delta\equiv \delta_\epsilon+\delta_\eta$, and   $(\epsilon^i,\epsilon_i)$ and $(\eta^i,\eta_i)$ parametrize the Poincar\'e and conformal supersymmetry transformations. $\mc F_{ab}$ is the superconformal covariant field strength (see equation (20.77) in~\cite{freedman12}) and 
	\begin{align}
	 \mc D_\mu X = (\p_\mu - b_\mu - iA^R_\mu) X - \frac{1}{2} \ov\psi_\mu^i \Om_i\,
	\end{align}
	 is the superconformal covariant derivative acting on the scalar field $X$. 
	In order to preserve the gauge choice (\ref{fixmass})(\ref{fixS}), we must accompany  the  Poincar\'e supersymmetry transformations  $\delta_\epsilon$ with a field dependent compensating conformal supersymmetry
	transformation $\delta_\eta$ with parameter\footnote{Since  (\ref{fixS}) preserves $\de X$, no other compensating transformation in required.}
	\be
  \eta_i = {i\over 2}   A\fs^R \ep_i - \frac{1}{2\mu} \lt(\frac{1}{4} \Ga^{ab} \mc F_{ab} \vep_{ij} + {Y_{ij}\over 2}\rt) \ep^j\,.
	\label{eta}
	\ee

	Different Poincar\'e supergravity theories     depend  on the choice of a second multiplet which gauge fixes the remaining $SU(2)_R$ symmetry. Three choices for this   compensating multiplet have been considered in the literature (see  \cite{deWit:1982na}): a non-linear multiplet, a hypermultiplet and a tensor multiplet. 
	We now demonstrate that the  $OSp(2|4)$-invariant $S^4$
is a supersymmetric background of the ${\cal N}=2$ Poincar\'e supergravity theory constructed  with a tensor multiplet (and not with the non-linear or hypermultiplet).
	
	Consider the  off-shell ${\cal N}=2$ Poincar\'e supergravity multiplet constructed by coupling a vector multiplet and a tensor multiplet to the Weyl multiplet.
An ${\cal N}=2$ tensor multiplet~\cite{deWit:1982na}
\begin{align}
	\text{bosonic}&:  L_{ij}, G, E_{mn}\cr
	\text{fermionic}&: \phi^i
	\end{align}
consists of a triplet of real scalars  $L_{ij}=L_{ji}$, a tensor gauge field $E_{mn}$, a complex scalar $G$ and a doublet of spinors $\phi^i$. 
The $SU(2)_R$ symmetry can be gauge fixed by setting
\be
L_{ij} = \tau_{1ij} \varphi\,, \label{breakSU2}
	\ee
	which breaks   $SU(2)_R$   down to $SO(2)_R$. The   supersymmetry transformation 
 \cite{parisLect} 
\be
		\de L_{ij} = \ov\ep_{(i} \phi_{j)} + \vep_{ik} \vep_{jl} \ov\ep^{(k}\phi^{l)} 
\ee	
implies that to   preserve (\ref{breakSU2}), we must accompany the Poincar\'e supersymmetry transformation $\delta_\epsilon$   with a compensating $SU(2)_R$ transformation $\delta_{SU(2)_R}(\Lm_{\;\,j}^k)$ with parameter\footnote{The parameter is determined only up to an $SO(2)_R$ transformation.} 
	\begin{align}
	 \Lm_{\;\,j}^k = -\tau_1^{km} {(\ov\ep_m \phi_j - \vep_{im} \vep_{jl} \ov\ep^i \phi^l)\over \varphi }\,.   \label{Lm}
	\end{align}
 In summary, this off-shell Poincar\'e supergravity multiplet constructed by gauge fixing a Weyl, vector and tensor multiplet completes the on-shell multiplet $e^a_m,\psi^i_m,A_m$ with   bosonic auxiliary fields   and fermionic auxiliary fields $\chi^i, \phi^i$. The Poincar\'e supersymmetry transformations in this ${\cal N}=2$ Poincar\'e supergravity theory are given by the following combination of superconformal transformations 
\be
  \delta_\epsilon+\delta_\eta +\delta_{SU(2)_R}(\Lm_{\;\,j}^k)
\ee
with $\eta$ in (\ref{eta}) and $\Lm_{\;\,j}^k$ in (\ref{Lm}).

In this ${\cal N}=2$ Poincar\'e supergravity theory the   supersymmetric backgrounds where the background values of all fermions vanish are solutions to the following equations 
	 \be
	 \left(\delta_\epsilon+\delta_\eta\right) \psi^i_m=0\qquad \left(\delta_\epsilon+\delta_\eta\right) \chi^i=0\qquad \left(\delta_\epsilon+\delta_\eta\right) \phi^i=0\,
	 \ee
	with $\eta$ in (\ref{eta}), since $\Lm_{\;\,j}^k=0$ vanish on bosonic backgrounds. 
 	The explicit form of these transformations are~\cite{deWit:1979ug}\cite{deWit:1979pq}\cite{deWit:1982na} (we use \cite{parisLect}) 
\begin{align}
\de \psi_m^i =& \lt(\p_m + \frac{1}{2}b_m + \frac{1}{4} \Ga^{ab} \om_{m ab} - \frac{1}{2} i A^R_m \rt)
			\ep^i + V_{m\;\,j}^{\;\;i} \ep^j - \frac{1}{16} \Ga^{ab} T_{ab}^- \vep^{ij} \ga_m \ep_j 
			- \ga_m \eta^i \nn\\
		\de \chi^i =& \frac{1}{2} D \ep^i + \frac{1}{6} \Ga^{ab} \lt[ -\frac{1}{4} \mc D \!\fs\, T_{ab}^- \vep^{ij} \ep_j
			- \widehat{R}_{ab} (U_j^{\;\,i}) \ep^j + i \widehat{R}_{ab}(T) \ep^i + \frac{1}{2} T_{ab}^- \vep^{ij} \eta_j
			\rt]\nn\\
			\de \phi^i =& \frac{1}{2} \slashed{D} L^{ij} \ep_j + \frac{1}{2} \vep^{ij} E\!\fs\, \ep_j - \frac{1}{2} G\ep^i + 2L^{ij}\eta_j\,,
			\label{BPSsugra}
\end{align}
with $\eta$ in (\ref{eta}). $ \mc D$ is superconformal covariant derivative and $\widehat{R}_{ab}(T)$ and $ \widehat{R}_{ab} (U_j^{\;\,i})$ are covariant curvatures for   $U(1)_R$ and $SU(2)_R$.
	
The $OSp(2|4)$- supersymmetric $S^4$ background is described by the following Killing spinor equations   (\ref{killSPH}) 
	\be
	\nabla_m \ep^i = \frac{i}{2r} \ga_m \tau_1^{ij} \ep_j\qquad \nabla_m \ep_i = \frac{i}{2r} \ga_m \tau_{1 ij} \ep^j\,.
	\label{fourkill}
	\ee
From 	(\ref{BPSsugra}) we find  that $S^4$ is a supersymmetric background of this supergravity theory with the following non-vanishing background fields turned on
\be
e^a_m=e^a_m|_{S^4}\qquad Y^{ij}=-{2i\mu\over r}\tau_1^{ij} \qquad Y_{ij}=-{2i\mu\over r}\tau_{1 ij}\qquad \text{other}=0\,. \label{S4back}
\ee
	With these background fields turned on    $\delta\psi_m^i$ realizes the $S^4$ Killing spinor equations (\ref{fourkill}), while $\delta\xi^i$ and $\delta\phi^i$ vanish identically.\footnote{A similar analysis for the Poincar\'e supergravity theories constructed with a compensating non-linear multiplet and hypermultiplet demonstrates that the background fields that yield the  $S^4$ Killing spinor equations are incompatible with the vanishing of the supersymmetry variations of the fermions in these multiplets. Therefore, $S^4$ is not a supersymmetric background of these supergravity theories.} The algebra of supergravity transformations when evaluated on the background (\ref{S4back}) realizes  the  $OSp(2|4)$ symmetry of $S^4$.

\section{The K\"{a}hler ambiguity Supergravity Counterterm}
\label{kahlersugrab}

In this section we construct the  ${\cal N}=2$ Poincar\'e supergravity invariant constructed out of the supergravity multiplet and  the $w=0$ chiral multiplets ${\cal A}_I$    (see below (\ref{spurii})) which when evaluated on the  $OSp(2|4)$-supersymmetric background (\ref{susbackg}) 
  realizes the K\"ahler ambiguity (\ref{kahler}). 

Our  approach is to   construct a   superconformal invariant constructed out of the Weyl multiplet, the compensating vector multiplet $\Phi$, the compensating tensor multiplet  and the   chiral multiplets ${\cal A}_I$, the supermultiplets to which the coordinates in the conformal manifold $\tau_I$ have been promoted.  This invariant, when evaluated on the Poincar\'e gauge fixing choice described in the previous section 
  yields an invariant in the associated ${\cal N}=2$ Poincar\'e supergravity theory.   We  first recall some facts about the   construction of   superconformal invariants.

Consider  an abstract chiral multiplet  (\ref{dC}) with $w=2$,  which we denote by $\hat{\cal A}$,  coupled to the Weyl multiplet (\ref{weylm}).   The following superconformal invariant can be  constructed  from  such a  chiral multiplet~\cite{deWit:1980tn} 
\be
I[\hat {\mc A}]=\int d^4x\sqrt{g}\left[\hat C(x) -\frac{1}{4} \hat A \left(T_{ab}^{+}\right)^2 + \text{fermions}  \right]\,,
\label{invaraiant}
\ee
where $\hat C$  and $\hat A$ denote  the top and bottom components of the multiplet $\hat{\cal A}$. The coupling of the chiral multiplet to the Weyl multiplet is responsible for the appearance of the terms after $\hat C$ in (\ref{invaraiant}). The product of two chiral multiplets with R-charge $w_1$ and $w_2$ yields another 
chiral multiplet of R-charge $w_1+w_2$. Therefore, superconformal invariants can be constructed from products of chiral multiplets with total R-charge $w=2$.

Consider  now the compensating vector multiplet that appears in the construction of ${\cal N}=2$ Poincar\'e supergravity, which we denote by $\Phi$. It is important to note that an ${\cal N}=2$   vector multiplet  is a chiral multiplet with $w=1$ subject to  reducibility constraints~\cite{deRoo:1980mm}, which express the last two components of the chiral multiplet in terms of the previous ones.
It is also known as a restricted chiral multiplet. The components of a chiral multiplet (\ref{dC}) are given in terms
of the fields in the abelian vector multiplet (\ref{vector})  by
	\begin{align}
		\lt. A\rt|_\Phi =&\; X \nn\\
		\lt. \Psi_i\rt|_\Phi =&\; \Om_i \nn\\
		\lt. B_{ij}\rt|_\Phi =&\; Y_{ij}   \nn\\
		\lt. F_{ab}^-\rt|_\Phi =&\; {\cal F}^-_{ab}  \nn\\
		\lt. \Lm_i\rt|_\Phi =&\; -\vep_{ij} \slashed{D} \Om^j \nn\\
		\lt. C\rt|_\Phi =&\; -2D_aD^a\bar X - \frac{1}{2} {\cal F}_{ab}^+ T^{ab+}   - 3\bar\chi_i \Om^i
		\label{phi}
	\end{align}
	where ${\cal F}_{ab}$ is the superconformal covariant field strength. Expressing a vector multiplet as a $w=1$ chiral multiplet provides a way of constructing a superconformal invariant out of $\Phi$ using (\ref{invaraiant}). 
	
	We will now construct the supergravity counterterm that realizes the K\"ahler ambiguity by writing down a supergravity invariant constructed out of a composite chiral multiplet with $w=2$. From    the compensating vector multiplet $\Phi$,   which has $w=1$, it is possible to construct two chiral multiplets with $w=2$ 
\be
		\mathds{T}(\log \bar\Phi) \quad \mbox{and} \quad \Phi^2\,,
	\ee
where the first is the so called 	 non-linear kinetic multiplet \cite{deWit:1980lyi}\cite{Butter:2013lta}.\footnote{Given an anti-chiral multiplet $\bar\cA$ with $w=0$, the corresponding kinetic multiplet $\mathds{T}(\bar{\cA})$, which has $w=2$, is  defined as: $\mathds{T}(\bar{\cA}) \propto \bar D^4 \bar{\cA}$, where $\bar D^4$ involves all four anti-chiral covariant superspace derivatives.} 
Given a chiral multiplet $\cA$ with bottom component $A$ and   R-charge $w$, the multiplet $\log \cA$ is a chiral multiplet whose bottom component, namely $\log A$, transforms inhomogeneously under dilatations but  its higher components in the multiplet (in particular the top component) transform as if they belonged to a chiral multiplet with $w=0$ \cite{Butter:2013lta}. 
The usefulness of this multiplet comes from the fact that the top component of  a chiral multiplet with $w=0$ is the bottom component of an anti-chiral multiplet with $w=2$, and in particular a superconformal primary (i.e. invariant under $S$-supersymmetry). Taking the CPT conjugate we find that the top component of $\log\bar\Phi$ is a chiral primary with $w=2$. Therefore we can build a chiral multiplet with $w=2$ by applying the $Q$-supersymmetry generators on the top component of $\log\bar\Phi$ and this multiplet is precisely $\dsT(\log\bar\Phi)$ \cite{Butter:2013lta}.

The supergravity counterterm responsible for the K\"ahler ambiguity is the superconformal invariant (\ref{invaraiant}) constructed from the $w=2$ composite chiral multiplet 
 \be
 {\cal F}({\cal A}_I) \dsT(\log\bar\Phi)\,,
 \ee
 where $\mc F$ is an arbitrary holomorphic function of the  $w=0$ chiral multiplets  ${\cal A}_I$  describing the coordinates in the conformal manifold.\footnote{An analogous $w=2$ chiral multiplet constructed out of $\Phi^2$, namely $\cF(\cA_I) \Phi^2$ can be used to construct another counterterm but when evaluated on the $OSp(2|4)$ invariant background (\ref{susbackg}) the invariant becomes: $I[\cF(\cA_I) \Phi^2] = 32\pi^2 \mu^2r^2$. In the original version of this paper this term was proposed as the counterterm responsible for the K\"ahler ambiguity but due to the quadratic scale dependence, this  counterterm   renormalizes the quadratic divergences of $\log Z_{S^4}$. Yet another natural guess for the K\"ahler counterterm can be  constructed from the  $w=2$ chiral multiplet $W^{ab}W_{ab}  {\cal F}({\cal A}_I)$, where $W_{ab}$ is a chiral multiplet that encodes the covariant  Weyl multiplet (\ref{weylm}). However, upon evaluating these terms on the background (\ref{susbackg}) they all vanish, as   these terms involve the Weyl tensor, which vanishes on $S^4$. Supergravity  couplings  involving $W^2$ have been considered in the literature~\cite{LopesCardoso:1998wt}. For other higher derivative invariants in ${\cal N}=2$ supergravity see e.g.~\cite{Butter:2010jm}\cite{deWit:2010za}\cite{Butter:2013lta}\cite{Kuzenko:2013gva}\cite{Butter:2014iwa}.}
 The associated  ${\cal N}=2$ Poincar\'e supergravity invariant is 
\be
I\left[{\cal F}({\cal A}_I) \dsT(\log\bar\Phi)\right]\,.
\label{invpoin}
\ee

We will now evaluate this invariant in the $OSp(2|4)$ invariant background field configuration:
\begin{align}
\text{Weyl:}\,& e^a_m=e^a_m|_{S^4}\nn\\
\text{vector:}\,& \lt. A\rt|_\Phi = X=\mu\,,\qquad \lt. B_{ij}\rt|_\Phi= -{2i\mu\over r}\tau_{1ij}\,, \qquad  \lt. C\rt|_\Phi ={4\mu\over r^2}\nn\\
\text{chiral:}\,&  \lt.  A\rt|_{{\cal F}({\cal A}_I)}={\cal F}(\tau_I)\,,
\label{susbackg}
\end{align}
Since the only field with nonzero expectation value in the Weyl multiplet is the vierbein, the only term from (\ref{invaraiant}) that survives in this background is the top component of the product chiral multiplet ${\cal F}({\cal A}_I) \dsT(\log\bar\Phi)$. The product of two chiral multiplets with bosonic components $(A,B_{ij},F_{ab}^-,C)$ and $(a,b_{ij},f_{ab}^-,c)$ yields a new chiral multiplet
with bosonic components (setting all fermions to zero, as they vanish in the $OSp(2|4)$-invariant background (\ref{susbackg}))
\be
\left(Aa, A b_{ij}+aB_{ij}, Af_{ab}^-+aF_{ab}^-,Ac+aC-{1\over 2}\varepsilon^{ik}\varepsilon^{jl}B_{ij}b_{kl}+F_{ab}^-f^{-ab}\right)\,. \label{prodChiral}
\ee
We need the bosonic components of the chiral multiples $\cF(\cA_I)$ and $\dsT(\log\bar\Phi)$ to compute the top component of their product. Since only the bottom component of $\cF(\cA_I)$ is nonzero in the background (\ref{susbackg}), to compute the top component of the product $\cF(\cA_I) \dsT(\log\bar\Phi)$ according to (\ref{prodChiral}) we only need to know the top component of $\dsT(\log\bar\Phi)$. The components of $\dsT(\log\bar\Phi)$ were computed in terms of the components of $\bar\Phi$ in \cite{Butter:2013lta}. Using their expressions, in the background (\ref{susbackg}), the top component becomes:
\begin{align}
	C|_{\dsT(\log\bar\Phi)} =&\; 4 D^2 D^2 \log\bar\mu - 8 \cR^{ab} D_a D_b \log\bar\mu + \frac{8}{3} \cR D^2 \log\bar\mu + \frac{2}{3} D^2 \cR - 2 \cR^{ab} \cR_{ab} + \frac{2}{3} \cR^2\,, \label{topT}
\end{align}
where $D_a$ is space-time covariant derivative, $D^2 \equiv D_aD^a$, $\cR_{ab}$ is the Ricci curvature on the sphere, and $\cR$ is the scalar curvature. Since $\log\bar\mu$ and $\cR$ are covariantly constant scalars all the terms in (\ref{topT}) with derivatives vanish and after substituting the values for $\cR_{ab}$ and $\cR$, (\ref{topT}) becomes:
\be
	C|_{\dsT(\log\bar\Phi)} = \frac{24}{r^4}\,, \label{topTev}
\ee
and finally
\be
	C|_{\cF(\cA_I) \dsT(\log\bar\Phi)} = \cF(\tau_I)\frac{24}{r^4}\,.
\ee
Thus, we find that the invariant (\ref{invpoin}) is:
\be 
	I\left[{\cal F}({\cal A}_I) \dsT(\log\bar\Phi)\right] = \int_{S^4} \sqrt{g}\, \cF(\tau_I)\frac{24}{r^4} = 64\pi^2 \cF(\tau_I)\,.
\ee

Therefore, the marginal supergravity counterterm
\be
{1\over 768 \pi^2}\left( I[\cF(\cA_I) \dsT(\log\bar\Phi)]+I[\bar\cF(\bar\cA_I) \dsT(\log\Phi)] \right)
\ee
is responsible for the K\"ahler ambiguity (\ref{kahler}) in the four sphere partition function of four dimensional ${\cal N}=2$ SCFTs
\be
Z_{S^4}\simeq Z_{S^4} e^{{1\over 12}\left({{\cal F}(\tau)+\bar{\cal F}(\bar \tau)}\right)}\,.
\ee
 	This provides a microscopic realization of K\"ahler ambiguities in these SCFTs.
	
	\section*{Acknowledgements}

We would like to thank D. Butter, B. de Wit, E. Gerchkovitz, Z. Komargodski, and  A. Van Proeyen for useful discussions.     Research at Perimeter Institute is
supported by the Government of Canada through Industry Canada and by the
Province of Ontario through the Ministry of Economic Development and
Innovation. J.G. also acknowledges further support from an NSERC
Discovery Grant and from an ERA grant by the Province of Ontario.

	\vfill\eject
\bibliography{refs}
	
\end{document}